\documentclass[10pt,english,aps,amssymb,prl,twocolumn, showpacs]{revtex4-1}
\pdfoutput=1
\usepackage[T1]{fontenc}
\usepackage[latin9]{inputenc}
\usepackage{amsmath}
\usepackage{graphicx}
\usepackage{amssymb}
\usepackage{esint}
\usepackage{esint}
\usepackage{verbatim}
\usepackage{mathtools}
\usepackage{wrapfig}
\usepackage{bbold}
\usepackage{amsfonts}

\usepackage{bm}
\usepackage{bbm}

\usepackage[top=2cm, bottom=2cm, left=2cm, right=2cm]{geometry}

\makeatletter
\@ifundefined{textcolor}{}
{% \definecolor{BLACK}{gray}{0}
 \definecolor{WHITE}{gray}{1}
 \definecolor{RED}{rgb}{1,0,0}
 \definecolor{GREEN}{rgb}{0,1,0}
 \definecolor{BLUE}{rgb}{0,0,1}
 \definecolor{CYAN}{cmyk}{1,0,0,0}
 \definecolor{MAGENTA}{cmyk}{0,1,0,0}
 \definecolor{YELLOW}{cmyk}{0,0,1,0}
 }

\renewcommand{\vr}[1]{{\mathbf #1}}
\renewcommand{\bm}{\vec}
\newcommand{\intt}{\int\!}

\renewcommand{\phi}{\varphi}
\renewcommand{\epsilon}{\varepsilon}

\renewcommand{\vec}[1]{{\mathbf #1}}

\usepackage{babel}

\begin{document}

\title {Topological superconductivity and high Chern numbers in 2D ferromagnetic Shiba lattices }
\author{Joel R\"ontynen}
\author{Teemu Ojanen}
\email[Correspondence to ]{teemuo@boojum.hut.fi}
\affiliation{O. V. Lounasmaa Laboratory (LTL), Aalto University, P.~O.~Box 15100,
FI-00076 AALTO, Finland }
\date{\today}
\begin{abstract}
Inspired by the recent experimental observation of topological superconductivity in ferromagnetic chains, we consider a dilute 2D lattice of magnetic atoms deposited on top of  a superconducting surface with a Rashba spin-orbit coupling. We show that the studied system supports a generalization of $p_x+ip_y$ superconductivity and that its topological phase diagram contains Chern numbers higher than $\xi/a$ $(\gg1)$, where $\xi$ is the superconducting coherence length and $a$ is the distance between the magnetic atoms. The signatures of nontrivial topology can be observed by STM spectroscopy in finite-size islands.   
       
\end{abstract}
\maketitle
\bigskip{}

\emph{Introduction}--The recent experiment reporting signatures of topological superconductivity in magnetic chains, consisting of arrays of magnetic atoms on top of a superconductor, has opened up remarkable chapter in the pursuit of novel topological phases of matter \cite{np2}.  Promising signatures of topological superconductivity and Majorana bound states have previously been reported in nanowire setups \cite{oreg, lutchyn, mourik, das}, although the later analysis of the zero-bias peak attributed to Majorana states has revealed a number of alternative explanations. The groundbreaking experiment in magnetic chains directly demonstrated that the midgap states are localized at the ends of the chain, corroborating the topological character of these states. These developments are important since realization of topological superconductivity in 1D networks \cite{alicea2, li} would open up route towards topological quantum computation \cite{nayak}.  

Motivated by the recent experiment and anticipating future developments, we consider a 2D lattice of ferromagnetic magnetic moments on a 2D superconducting surface with a Rashba spin-orbit coupling. Magnetic moments bind Yu-Shiba-Rusinov subgap states \cite{yu, shiba, rusinov, salkola, yaz}  with wavefunctions decaying  as $\frac{e^{-r/\xi}}{r^{1/2 } }$. Therefore the Shiba states have strong overlap with a large number of neighbouring sites  when $a<\xi$, where $\xi$ is the superconducting coherence length and $a$ is the lattice constant of the magnetic atoms. In the regime where the direct overlap of the orbitals of the magnetic atoms is negligible, the hybridization of the Shiba states still enable a subgap band formation. Following the treatments in Refs.~\cite{pientka2, pientka3, bry}, we derive an effective long-range 2D hopping model and study its topological properties in the deep-dilute impurity regime. The characteristic energy scales of the system are the isolated Shiba energy $\epsilon_0$ and the hybridization energy $\frac{\Delta}{(k_Fa)^{1/2}}$ of two impurities. We study the topological phase diagram as a function of these parameters by evaluating Chern numbers classifying the phases. In the physically relevant circumstances the distance between adjacent magnetic moments satisfy $\xi/a\sim10-10^3$, so the effective Hamiltonian describing the Shiba lattice includes long hoppings between $\mathcal{O}(\xi/a)$ nearest neighbours before the exponential suppression cuts them off. The detailed properties of the 1D long-range Shiba models \cite{pientka2, pientka3, bry, heimes, heimes2, ront, west, reis}  are known to  have important differences compared to the short-range toy models \cite{choy, np, vazifeh, poyh}. We show that the competition between a large number of long-range hopping terms gives rise to a complicated Chern number hierarchy. The studied system generally supports phases with high Chern numbers of the order of $\xi/a$ which leads to significantly richer phase diagram compared to short-range toy models \cite{nakosai}. Nonvanishing Chern numbers indicate the existence of gapless edge states that could be probed in STM experiments. We will show that the Local Density of States (LDOS) in finite-size systems exhibits signatures of the edge states, providing smoking-gun evidence of the bulk topological order.   
\begin{figure}
%\centering
\includegraphics[width=0.7\columnwidth]{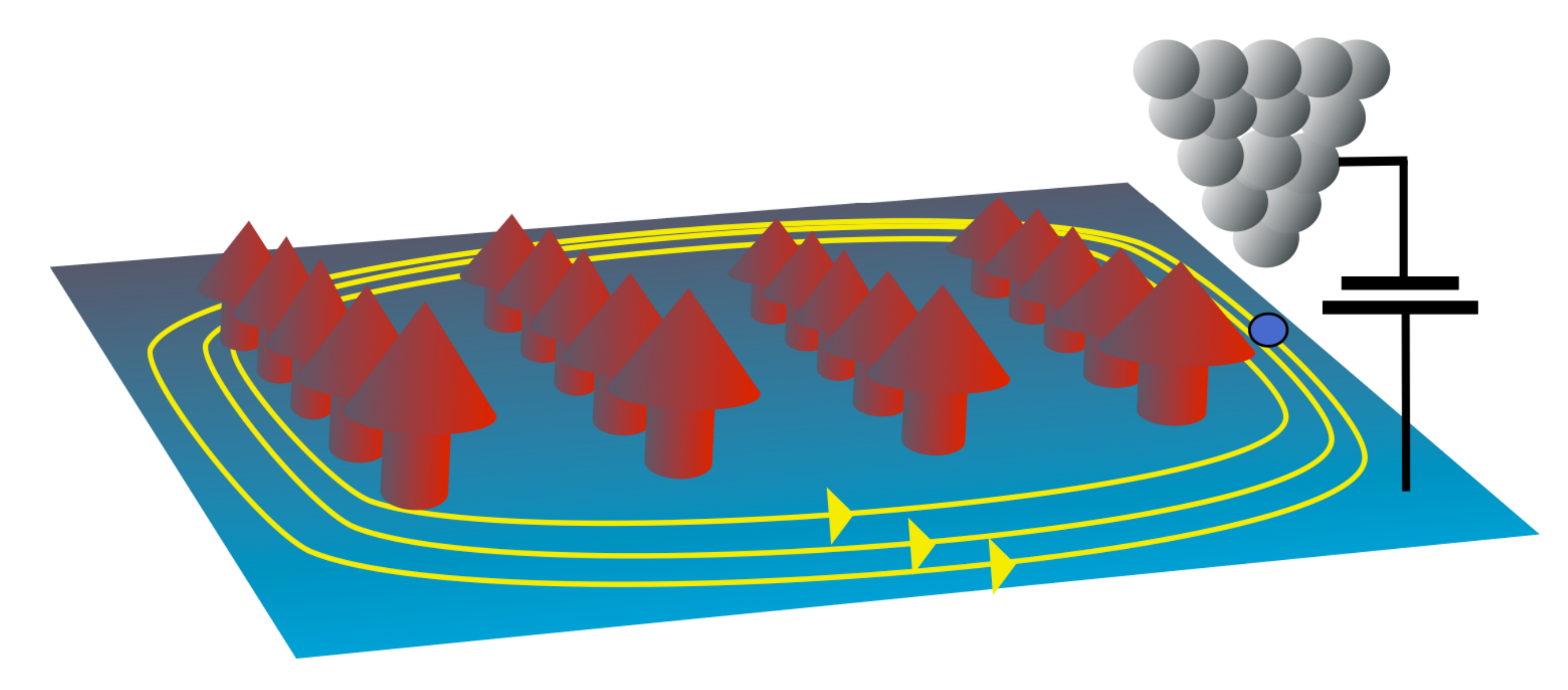}
\caption{ Array of magnetic impurities on an $s$-wave superconductor form a 2D Shiba lattice. This system supports a generalized $p_x+ip_y$ superconductivity with high Chern numbers. The subgap density of states due to the gapless edge states can be probed by STM spectroscopy. }\label{fig1}
\end{figure}

\emph{Model of ferromagnetic Shiba lattices}-- We begin by outlining the derivation of a low-energy model describing the subgap spectrum of a 2D s-wave superconductor with an array of magnetic impurities arranged in a 2D lattice such as the one in Fig~\ref{fig1}. The model is valid for general 2D lattice geometries, though later we consider a square lattice. The derivation proceeds similarly to the one presented in Ref.~\cite{pientka2} for a helical Shiba chain and that of the 1D ferromagnetic chain in Ref.~\cite{bry}. Therefore the details are relegated to the supplement. 

The Bogoliubov-de Gennes (BdG) Hamiltonian describing the system is $\mathcal{H}=\mathcal{H}^{(\rm bulk)}+\mathcal{H}^{(\rm imp)}$, consisting of two parts
\begin{align}\label{Hbulk}
&\mathcal{H}^{(\rm bulk)} = \tau_z \big[ \xi_\vec{k} + \alpha_R( k_y\sigma_x - k_x\sigma_y )\big] + \Delta\tau_x,\nonumber\\
&\mathcal{H}^{(\rm imp)} = -J\sum_j \vec{S}_j\cdot\boldsymbol{\sigma}\delta(\vec{r}-\vec{r}_j),
\end{align}
where $\mathcal{H}^{(\rm bulk)}$ describes bulk electrons in a 2D system and $\mathcal{H}^{(\rm imp)}$ represents the contribution of the magnetic atoms forming a lattice. These expressions have been written in the Nambu spinor basis  $\hat{\Psi}=(\hat{\psi}_{\uparrow},\hat{\psi}_{\downarrow},\hat{\psi}_{\downarrow}^\dagger,-\hat{\psi}_{\uparrow}^\dagger)^T$ and the Pauli matrices $\boldsymbol{\tau}$ and $\boldsymbol{\sigma}$ describe the particle-hole and the spin degree of freedom. In the above equations $\xi_\vec{k}$ is the kinetic energy, $\alpha_R$ is the Rashba spin-orbit coupling and $\Delta$ is the superconducting pairing in the substrate. In the absence of superconductivity the 2D bulk has two Rashba-split Fermi surfaces with distinct Fermi momenta $k_F^\pm = k_F \big( \sqrt{1 + \lambda\displaystyle{^2}} \mp \lambda \big)$ and the densities of states at the Fermi level $\mathcal{N}_{\pm} = \mathcal{N} \Big[ 1 \mp \lambda\big/\sqrt{1 + \lambda\displaystyle{^2}} \Big]$. Here we have defined a dimensionless spin-orbit strength $\lambda = \alpha_R/(\hbar v_F)$, the density of states $\mathcal{N}$ and the Fermi velocity $v_F$ in the absence of the Rashba coupling. Magnetic atoms, located at positions $\vec{r}_j$, are characterized by their spin $\boldsymbol{S}_j$ and coupling to the bulk electrons with the exchange coupling $J$.  Motivated by Ref.~\cite{np2}, we concentrate on the ferromagnetic ordering  where all $\boldsymbol{S}_j$ are perpendicular to the surface.  After a number of steps outlined in the supplement \cite{supp}, the BdG  eigenvalue problem  $\mathcal{H}\Psi = E \Psi$ leads to the relation 
\begin{equation} \label{TwoBandEnd}
(\boldsymbol{S}_i \cdot \sigma-J_E(0))\Psi(\mathbf{r}_i)=-\sum_{j\neq i}J_E(\mathbf{r}_i-\mathbf{r}_j)\Psi(\mathbf{r}_j), 
\end{equation}
where $J_E(\vec{r}) = JS \intt \frac{d\vec{k}}{(2\pi)^2} e^{i\vec{k}\cdot\vec{r}} \big[ E - \mathcal{H}^{(\rm bulk)}_\vec{k} \big]^{-1}$ and $S=|\boldsymbol{S}_j|$.  Relation (\ref{TwoBandEnd}) provides a closed set of equations for the spinor at the impurity positions. 

Due to the doubling of the degrees of freedom in the BdG formulation, a single magnetic atom will give rise to two subgap bound states with energies $\pm\epsilon_0=\pm\Delta\frac{1-\alpha^2}{1+\alpha^2}$ where  $\alpha=\pi\mathcal{N}JS$ is the dimensionless impurity strength.  As discussed in Refs.~\cite{pientka2, west,bry}, for a deep-dilute impurity arrangement satisfying $\alpha\approx1$ and $\frac{1}{(k_Fa)^{1/2}}\ll1$ we can accurately study the Shiba bands in the two-component basis $\Psi_j '(\vec{r}_j)\equiv \Psi_j '=\begin{pmatrix} u(\vec{r}_j) & v(\vec{r}_j)\end{pmatrix}^T$ of decoupled impurity states at site $\vec{r}_j$. Here $u(\vec{r}_j)$ and $v(\vec{r}_j)$  are the eigenstates to the single-impurity problem with energies $\epsilon_0\approx\Delta(1-\alpha)$ and $-\epsilon_0$.  

Projecting Eq.~(\ref{TwoBandEnd}) to the two basis states we obtain a reduced problem $H \Psi' = E\Psi'$
where
 \begin{equation}\label{H}
H_{ij}=\begin{pmatrix} h_{ij} & \Delta_{ij} \\ (\Delta_{ij})^\dagger & -h_{ij} \end{pmatrix}. 
\end{equation}
The effective BdG Hamiltonian (\ref{H}) is determined by the matrices
\begin{equation*}
\begin{split}
h_{ij} &= 
\Bigg\{\!\!\begin{array}{cl}
\epsilon_0 & i=j
\\
\displaystyle
-\frac{\Delta^2}{2} \big[ \tilde{I}_1^-(r_{ij}) + \tilde{I}_1^+(r_{ij}) \big] & i\neq j
\end{array} 
\\
\Delta_{ij} &= 
\Bigg\{\!\!\begin{array}{cl}
0 & i=j
\\
\displaystyle
\frac{\Delta}{2}  \big[ \tilde{I}_0^+(r_{ij}) - \tilde{I}_0^-(r_{ij}) \big] \frac{x_{ij} - iy_{ij}}{r_{ij}} & i\neq j 
\end{array} .
\end{split}
\end{equation*}
In the above expression $r_{ij}=|\vec{r}_i-\vec{r}_j|$, and $x_{ij}$ and $y_{ij}$ are components of $\vec{r}_i-\vec{r}_j\equiv(x_{ij},y_{ij})$. The hopping elements are expressed in terms of the functions
\begin{equation*}
\begin{split}
&\tilde{I}_0^\pm(r) = \frac{\mathcal{N}_\pm}{\mathcal{N}}\, \Re \Big[ iJ_{1}\big(k_F^\pm r +ir/\xi\big) + H_{-1}\big(k_F^\pm r +ir/\xi\big) \Big] \\
&\tilde{I}_1^\pm(r) = \frac{\mathcal{N}_\pm}{\mathcal{N}} \frac{1}{\Delta}
\Re \Big[ J_0\big( k_F^\pm r +ir/\xi \big) +iH_0\big( k_F^\pm r +ir/\xi \big) \Big],
\end{split}
\end{equation*}
where $J_n$ and $H_n$ denote the Bessel and Struve functions of order $n$, $\xi=\frac{v_F\sqrt{1+\lambda^2}}{\Delta}$ is the Rashba modified coherence length and $\Re$ stands for the real part of the expression on its right side. The effective model (\ref{H}) is valid for deep-lying energy states $E\ll\Delta$, with corrections proportional to $(\frac{E}{\Delta})^2$.

The key ingredients of the low-energy description (\ref{H}) are the coexistence of superconductivity, ferromagnetic ordering and the Rashba coupling, all of which have been demonstrated in the recent experiment \cite{np2}. To appreciate the crucial role of the spin-orbit coupling we note that $\Delta_{ij}$ vanishes when $\alpha_R=0$. The pairing function has the form $\Delta_{ij}=\Delta(x_{ij}-iy_{ij})f(r_{ij})$, where $f(r_{ij})= [\tilde{I}_0^+(r_{ij}) - \tilde{I}_0^-(r_{ij}) ]/2$. This indicates that the low-energy description (\ref{H}) has an odd-pairing symmetry $\Delta_{ij}=-\Delta_{ji}$ that generalizes the $p_x+ip_y$-type pairing. An ordinary chiral $p$-wave pairing would result if $f(r_{ij})$ was non-vanishing only for the nearest-neighbour hopping on a square lattice. However, in the model (\ref{H}) the normal hopping and the pairing function decay as $f(r)\propto\frac{e^{-r/\xi}}{r^{1/2}}$ \cite{supp}, so in the physically relevant case $\xi/a \sim10-10^3$ the model includes non-negligible hopping between dozens or hundreds of nearest neighbours. The relation between the model (\ref{H}) and a 2D chiral $p$-wave superconductor is similar to the relation between the long-range 1D Shiba models and Kitaev's toy model \cite{kitaev1}. Since a chiral $p$-wave pairing is the prototype of 2D topological superconductivity, it is natural to expect that the model (\ref{H}) also supports topologically nontrivial phases.  Below we will discuss how the long-range hopping has a dramatic impact on the topological properties and leads to remarkably complex phase diagrams.

\emph{Topological properties}-- Topological properties of the model (\ref{H}) on a square lattice are conveniently studied in momentum space. Defining Fourier transforms 
\begin{align}
d_x(\vec{k})&=\Re\sum_j \Delta_{ij}e^{ik_xx_{ij}+ik_yy_{ij}}\nonumber\\
d_y(\vec{k})&=\Im\sum_j\Delta_{ij}e^{ik_xx_{ij}+ik_yy_{ij}}\nonumber\\
d_z(\vec{k})&=\sum_jh_{ij}e^{ik_xx_{ij}+ik_yy_{ij}},\nonumber
\end{align}
the Hamiltonian in momentum space is expressed as $H(\vec{k})=\vec{d}(\vec{k})\cdot\boldsymbol{\sigma}$ with energies $E(k)=\pm|\vec{d}|$. The components of $\vec{d}=(d_x,d_y,d_z)$ do not allow a representation in terms of the elementary functions. The topological phase diagram of the studied model, which belongs to the Altland-Zirnbauer symmetry class D \cite{schnyder}, is revealed by evaluating the Chern number
\begin{equation}\label{chern}
C=\frac{1}{4\pi}\int_{\rm{BZ}}d^2\vec{k}\,\frac{\vec{d}\cdot\partial_{k_x}\vec{d}\times\partial_{k_y}\vec{d}}{|\vec{d}|^3},
\end{equation}
where the integration is performed over the Brillouin zone $k_x,k_y\in [-\frac{\pi}{a},\frac{\pi}{a}]$. The Chern number takes integer values and describes how many times the vector $\hat{\vec{d}}=\vec{d}/|\vec{d}|$ wraps around the unit sphere. Below we compute Chern numbers as a function of the relevant parameters $\epsilon_0$ and $k_Fa$.  The bulk-boundary correspondence implies that topological states with Chern number $C=q$ support $|q|$ branches of chiral gapless modes localized near the edge. The sign of $C$ determines the chirality of the edge modes. In non-superconducting systems $C$ determines a quantized Hall conductance whereas in superconducting systems only the thermal Hall conductance is quantized and the edge states are propagating Majorana modes \cite{volovik}.  

To understand qualitative features of the phase diagram, it is important to consider the connection between the long-range hopping and the Chern number. Intuitively this can be understood by noting that the $n$th hopping in $x$ and $y$ direction gives rise to such terms in $d_i$ as $\cos\,( nk_{x/y}a), \sin\, (nk_{x/y}a)$ that oscillate more rapidly with increasing $n$. Thus, the number of times $\hat{\vec{d}}$ may cover the unit sphere will generally increase with $n$. Employing the asymptotic approximations for the Bessel and Struve function \cite{supp},  one can see that the $n$th hopping terms decay as  $\frac{|\Delta|}{(k_Fa)^{1/2}}\frac{e^{-an/\xi}}{n^{1/2}}$, so the decay is very slow for the hopping range $n<\xi/a$. In addition to the monotonic decay, the $n$th hopping terms oscillate rapidly with wave vectors  $nk_F^{\pm}a$ so the phase diagram results from an effective  competition of roughly $\mathcal{O}(\xi/a)$ different hopping terms. In a recent study of toy models of two-band Chern insulators it was discussed how models with hopping range $n$ may give rise to Chern numbers scaling between $n$ and $n^2$ depending on the details of the model \cite{udagawa}. Remarkably, the model (\ref{H}) provides a concrete physical realization of a topological superconductor where Chern numbers are of the order of or larger than the effective hopping range $\xi/a$, as in the toy insulator models studied in Ref.~\cite{udagawa}.   

\begin{figure}
%\centering
\includegraphics[width=1.0\columnwidth]{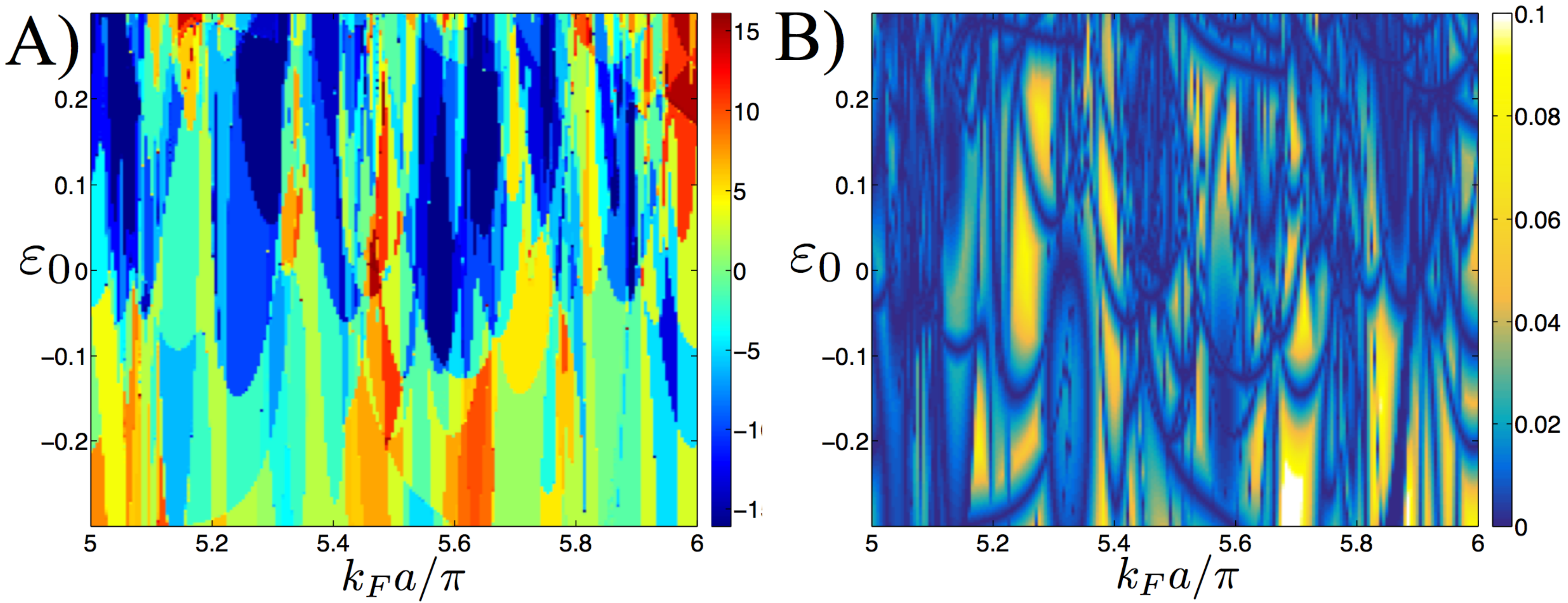}
\caption{ A) Chern number phase diagram for $\xi/a=10$ and $\lambda=0.05$. B) Minimum of the positive energy branch $\min_k E(k)$  in units of $\Delta$ for the same parameters. Different phases are separated by energy gap closing.}\label{fig2}
\end{figure}
\begin{figure}
%\centering
\includegraphics[width=1.0\columnwidth]{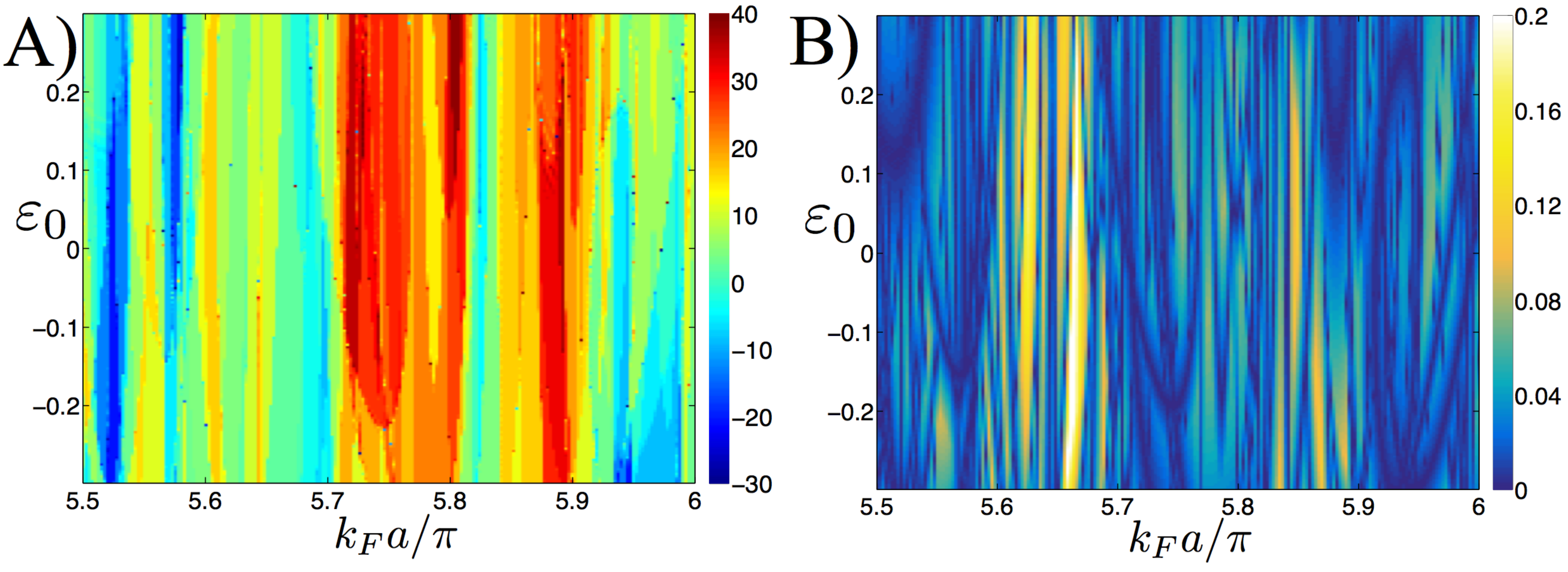}
\caption{ A) Chern number phase diagram for $\xi/a=30$ and $\lambda=0.05$. B)  Minimum of the positive energy branch $\min_k E(k)$ in units of $\Delta$ for the same parameters used in A).}\label{fig3}
\end{figure}
\begin{figure}
%\centering
\includegraphics[width=0.6\columnwidth]{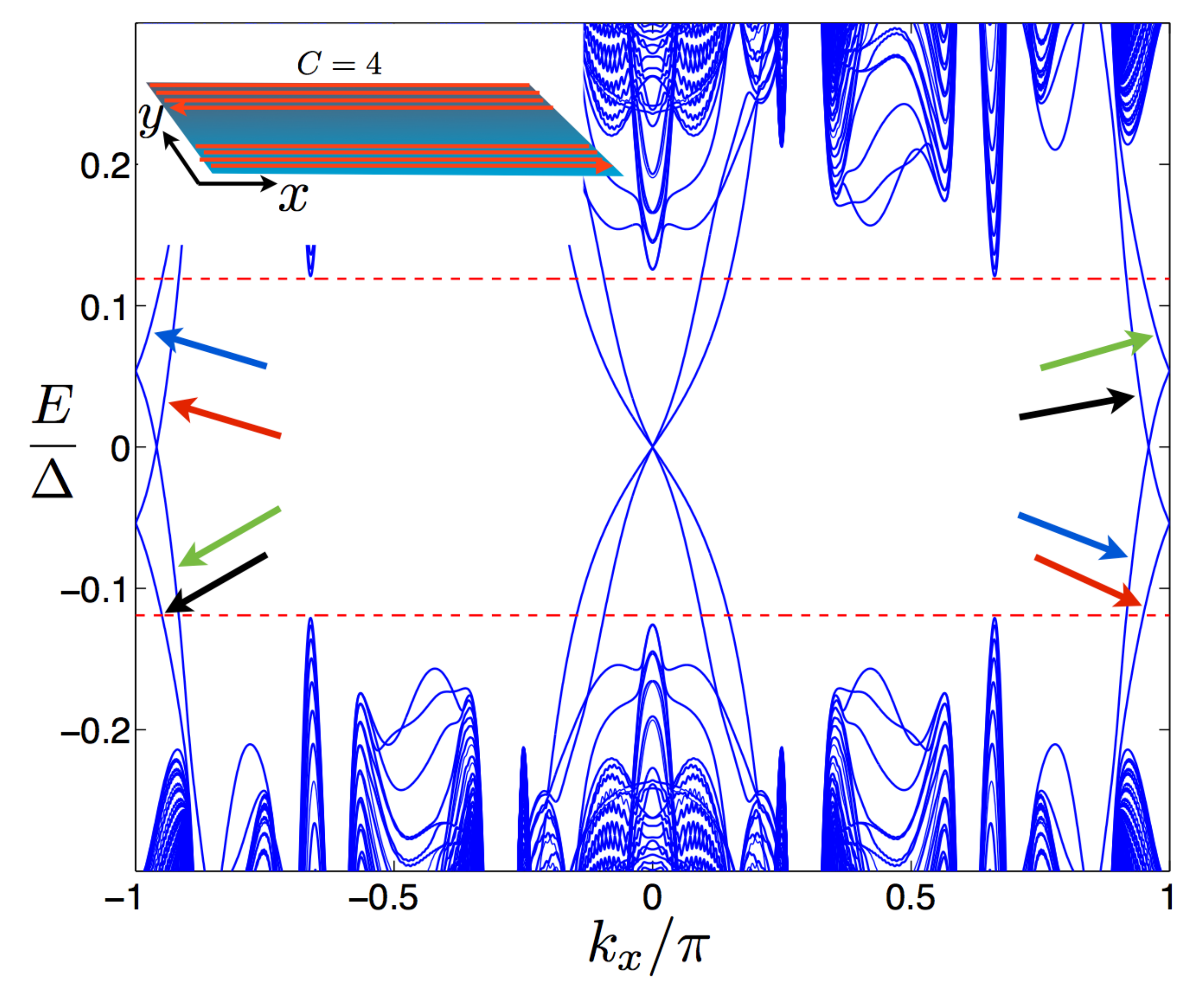}
\caption{Spectrum of an infinite strip as a function of momentum $k_x$ corresponding to Chern number $C=4$. Both edges support four chiral edge modes traversing the gap. Due to the periodicity of the Brillouin zone, the edge states close to $k_x=-\pi$ and $k_x=\pi$ describe the same set of states as indicated by the arrows. The dotted line marks the gap edge calculated for an infinite system. The figure corresponds to the case $\xi/a=10$,  $\epsilon_0/\Delta=-0.25$, $k_Fa/\pi=3.56$, $\lambda=0.05$ and the length in the $y$ direction is $L_y=200a$ }\label{fig4}
\end{figure}

In Figs.~\ref{fig2} and \ref{fig3} we have plotted topological phase diagrams as a function of the single-impurity energy $\epsilon_0$ and parameter $k_Fa$ controlling the hybridization of the impurity states. Different Chern numbers classify different phases that are separated by a closing of the energy gap determined by the condition $\min_k E(k)=0$. One can clearly see that for spin-orbit strength $\lambda=0.05$ corresponding to a momentum splitting $|k_F^{\pm}-k_F|=0.05k_F$ give rise to a large number of different phases with high Chern numbers and topological energy gaps $E_{\mathrm{gap}}=2\min_kE(k)$ of the order of $0.1\Delta$. These energies are still within the validity regime of the low-energy description (\ref{H}). The number of different topological phases and the highest Chern numbers having non-negligible occurrence are of the order of $\xi/a$. In Fig.~\ref{fig4} we plot the spectrum of (\ref{H}) calculated in a strip geometry. Diagonalization in a semi-infinite system reveals the existence of the edge states dictated by the bulk-boundary correspondence. 

\emph{Observable consequences and discussion}-- Chern numbers classify different topological states but are challenging to access in superconductors. The topological edge modes support a quantized thermal conductance which is much more difficult to observe than the ordinary quantized Hall conductance. A great advantage of Shiba systems is that they can be probed locally by STM spectroscopy. Signatures of Majorana wavefunctions localized at the ends of magnetic chains were recently observed. Analogously, the signatures of 2D topological order could be detected through STM spectroscopy where the edge states  could be observed in the LDOS of finite Shiba arrays as indicated in Fig.~\ref{fig1}.   
    
\begin{figure}
%\centering
\includegraphics[width=0.6\columnwidth]{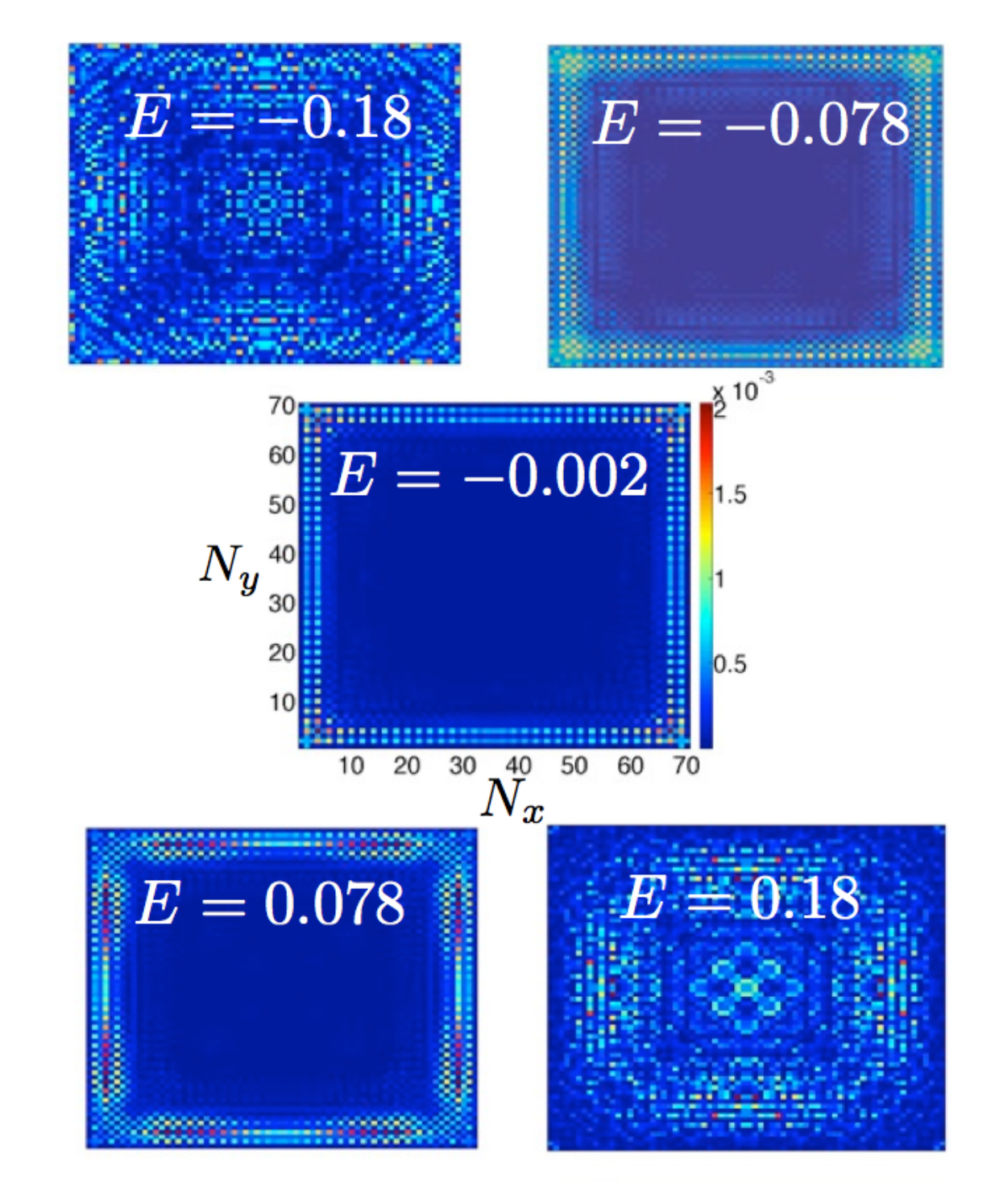}
\caption{Local density of states in a finite $70\times70$ lattice (in arbitrary units) corresponding to a bulk state with $C=3$. All the energies are in the units of $\Delta$ and correspond to tunnelling voltages $V=E/e$. Near the center of the gap the LDOS is suppressed in the bulk but enhanced on the edges due to the topological edge states. The figure corresponds to the case $\xi/a=10$,  $\epsilon_0/\Delta=-0.22$, $k_Fa/\pi=4.9$, $\lambda=0.05$}\label{fig5}
\end{figure}

Diagonalization (\ref{H}) in a finite square lattice enables us to evaluate the LDOS defined by $N(\vec{r},E)=\sum_n|u_n(\vec{r})|^2\delta(E-E_n)+|v_n(\vec{r})|^2\delta(E+E_n)$. Here $u_n(\vec{r})$ and $v_n(\vec{r})$ are the particle and hole components of the eigenstate with energy $E_n$. In the absence of magnetic atoms the system is in the trivial state and $N(\vec{r},E)=0$ for $|E|<\Delta$. However, the topological edge modes of finite Shiba lattices with $C\neq0$ show up in the subgap LDOS. Away from the phase boundaries the bulk spectrum is always gapped, while the edge states traverse the gap. Therefore the LDOS near the center of the gap $E/\Delta\ll 1$ should reveal the existence of topological edge states. Furthermore, since the edge states are localized at the sample edge  we expect that the midgap LDOS $N(\vec{r},E)$ is peaked when the coordinate $\vec{r}$ is located near the boundary and suppressed in the bulk. As illustrated in Fig.~\ref{fig5}, even relatively small systems exhibit these important features. STM spectroscopy is not sensitive to the precise value of  Chern number of the state but can detect nonzero values through the subgap LDOS. The fact that the low-lying excitations are localized in the vicinity of the edges provides a strong evidence of the bulk topological order in the system.       

As discussed in Ref.~\cite{np2}, the next challenges after the confirmation of the topological superconductivity in 1D ferromagnetic chains include studies of topological properties of 2D islands observed in the experimental setup. In the experiment iron atoms are densely packed a few \AA ngstr\"om apart so that the atomic $d$ orbitals overlap directly and give rise to the ferromagnetic ordering. Our theory addresses the situation where the distance between magnetic moments is of the order of nanometers and the direct overlap of atoms is negligible. In this case the ferromagnetic ordering may be obtained due to the interplay of RKKY coupling,  Rashba coupling and crystal field splitting \cite{heimes2}. The key ingredients leading to the rich topological properties discovered in our work are the coexistence of superconductivity, the ferromagnetic ordering of adatoms and the Rashba coupling on the surface, all of which are confirmed in Ref.~\cite{np2}.  As discussed above, the energy gaps for various Chern number phases can reach a few multiples of $0.1\Delta$ corresponding to temperatures of the order of $1K$. Considering that Pb surfaces may give rise to a spin-orbit coupling comparable to the one assumed in our calculations and that STM signatures of the edge modes are observable already in small systems, chances of finding 2D topological superconductivity in ferromagnetic islands seems very promising.

\emph{Summary and outlook}-- Motivated by the recent experimental discovery of topological superconductivity in ferromagnetic chains, we studied a 2D ferromagnetic Shiba lattices. We reported that these systems support a generalized $p_x+ip_y$ superconductivity with a large number of phases and Chern numbers higher than $\xi/a\gg1$ where $\xi$ is the superconducting coherence length and $a$ is the Shiba lattice constant. As in the 1D case, the signatures of topological edge states can be observed by STM spectroscopy. A more systematic exploration of phase diagrams, the topological properties of different lattice geometries and lattice imperfections, and exploration of different scenarios to tune the topological properties are left for future studies.  

The authors acknowledge Alex Weststr\"om and Kim P\"oyh\"onen for illuminating discussions, the computational resources provided by Aalto Science-IT project, and the Academy of Finland for support.

\appendix
\numberwithin{equation}{subsection}

\widetext
\subsection*{Supplemental material - derivation of the effective Hamiltonian in 2D Shiba lattice}
The Bogoliubov-de Gennes Hamiltonian can be separated into a bulk and an impurity term, $\mathcal{H} = \mathcal{H}^{(\rm bulk)} + \mathcal{H}^{(\rm imp)}$, where the bulk Hamitonian,
\begin{equation*}
\begin{gathered}
\mathcal{H}_\mathbf{k}^{(\rm bulk)} = \tau_z \big[ \xi_\vr{k} \sigma_0 + \alpha_R( k_y\sigma_x - k_x\sigma_y )\big] + \Delta\tau_x\sigma_0 ,
\end{gathered}
\end{equation*}
while $\xi_\vr{k} = \frac{\hbar^2k^2}{2m} - \mu$ with the Fermi energy $\mu$, $\lambda$ is the Rashba SOC strength and $\Delta$ is the superconducting pairing amplitude. The impurity Hamiltonian,
\begin{equation*}
\begin{gathered}
\mathcal{H}^{(\rm imp)}(\vr{r}) = -J\sum_j \vr{S}_j\cdot\bm{\sigma}\,\delta(\vr{r}-\vr{r}_j),
\end{gathered}
\end{equation*}
describes an exchange interaction of strength $J$ with the magnetic impurity atoms having spin $\vec{S}_j$ located at position $\vr{r}_j$. 

The BdG equation $\mathcal{H}(\vr{r}) \Psi(\vr{r}) = E \Psi(\vr{r})$ yields
\begin{equation*}
\begin{gathered}
\big[ E - \mathcal{H}^{(\rm bulk)}(\vr{r}) \big] \Psi(\vr{r}) 
%= \mathcal{H}^{(\rm imp)}(\vr{r}) \psi(\vr{r})
= -J\sum_j \vr{S}_j\cdot\bm{\sigma}\,\delta(\vr{r}-\vr{r}_j) \Psi(\vr{r}_j).
\end{gathered}
\end{equation*}
We change to momentum space using the Fourier transform $\Psi(\vr{r}) = \intt \frac{d\vr{k}}{(2\pi)^2} e^{i\vr{k}\cdot\vr{r}} \Psi_\vr{k}$ and thus obtain
\begin{equation*}
\begin{gathered}
\big[ E - \mathcal{H}^{(\rm bulk)}_\vr{k} \big] \Psi_\vr{k} = -J\sum_j \vr{S}_j\cdot\bm{\sigma}\, e^{-i\vr{k}\cdot\vr{r}_j} \Psi(\vr{r}_j).
\end{gathered}
\end{equation*}
Solving for $\psi_\vr{k}$ and changing back to real space we have 
\begin{equation}
\begin{gathered}
\Psi(\vr{r}) = - \sum_j J_E(\vr{r}-\vr{r}_j)\, \hat{S}_j\cdot\bm{\sigma}\, \Psi(\vr{r}_j),
\label{impurityeq}\tag{A.1}
\end{gathered}
\end{equation}
where $S=|\vr{S}|$, $\hat{S}=\vr{S}/S$ and
\begin{equation*}
\begin{gathered}
J_E(\vr{r}) = JS\! \intt \frac{d\vr{k}}{(2\pi)^2} e^{i\vr{k}\cdot\vr{r}} \big[ E - \mathcal{H}^{(\rm bulk)}_\vr{k} \big]^{-1} .
\end{gathered}
\end{equation*}

The Rashba SOI lifts the spin degeneracy and gives rise to two helicity bands with dispersions $\xi_\pm = \xi_k \pm \alpha_R k$. Here we have suppressed the momentum index in the helical dispersions. The propagator  splits into two helical sectors, $\big[ E - \mathcal{H}^{(\rm bulk)}_\vr{k} \big]^{-1} = \frac{1}{2} \big(G_- + G_+\big)$, where
\begin{equation*}
\begin{gathered}
G_\pm = \frac{\big( E\tau_0 + \xi_\pm\tau_z + \Delta\tau_x \big)\big( \sigma_0 \pm \sin\varphi\,\sigma_x \mp \cos\varphi\,\sigma_y \big)}{E^2 - \xi_\pm^2 - \Delta^2}
\end{gathered}
\end{equation*}
and $\vr{k} = k(\cos\varphi,\sin\varphi)$.

\subsubsection*{Single impurity}
Let us first consider a single impurity at the origin with spin $\vr{S}$. We can set $\vr{r}=\vr{0}$ in Eq.\ (\ref{impurityeq}) and obtain
\begin{equation*}
\begin{gathered}
\big[\mathbbm{1} + J_E(\vr{0})\, \hat{S}\cdot\bm{\sigma}\big] \Psi(\vr{0}) = 0.
\end{gathered}
\end{equation*}
As we are considering the limit of deep impurities, we can perform the integrals in $J_E(\vr{0})$ assuming $|E|<\Delta$, yielding
\begin{equation}
\begin{gathered}
\Big[\mathbbm{1} - \frac{\alpha}{\sqrt{\Delta^2-E^2}} (E\tau_0 + \Delta\tau_x) \hat{S}\cdot\bm{\sigma}\Big] \Psi(\vr{0}) = 0,
\label{singleimp}\tag{A.2}
\end{gathered}
\end{equation}
where $\alpha = \pi JS\mathcal{N}$ is a dimensionless impurity strength and $\mathcal{N} = \frac{1}{2\pi}\frac{m}{\hbar^2}$ the density of states at the Fermi level in the absence of SOI. 

Eq.\ (\ref{singleimp}) has two solutions, $|\tau_x+\rangle|\uparrow\rangle$ and $|\tau_x-\rangle|\downarrow\rangle$ with eigenvalues $E = \Delta\frac{1-\alpha^2}{1+\alpha^2}$ and $E = -\Delta\frac{1-\alpha^2}{1+\alpha^2}$, respectively. Here $\tau_x|\tau_x\pm\rangle = \pm |\tau_x\pm\rangle$, $\,\hat{S}\cdot\bm{\sigma}|\uparrow\rangle = |\uparrow\rangle$ and $\hat{S}\cdot\bm{\sigma}|\downarrow\rangle = -|\downarrow\rangle$.

\subsubsection*{Impurity lattice}
In case of multiple impurities with positions $\vr{r}_i$ and spins $\vr{S}_i$, Eq.\ (\ref{impurityeq}) becomes
\begin{equation*}
\begin{gathered}
\big[\mathbbm{1} + J_E(\vr{0})\, \hat{S}_i\cdot\bm{\sigma}\big] \Psi(\vr{r}_i) = - \sum_{j\neq i} J_E(\vr{r}_i-\vr{r}_j)\, \hat{S}_j\cdot\bm{\sigma}\, \Psi(\vr{r}_j).
\end{gathered}
\end{equation*}
We consider the limit of deep impurities, $\alpha\approx 1$, so that the energy of an isolated impurity state $\epsilon_0$ lies close to the center of the superconducting gap $\epsilon_0\ll\Delta$. We also assume that the impurity separation $a$ is large enough for the impurity band to be well within the superconducting gap, $E\ll\Delta$. As explained in Refs~[14, 16, 28], with these assumptions we can linearise the LHS w.r.t.\ $E$ and $1-\alpha$ and evaluate the coupling term on the RHS for $E=0$ and $\alpha=1$:
\begin{equation}\label{J}\tag{A.3}
\begin{gathered}
\Big[\mathbbm{1} - \big(E/\Delta\,\tau_0 + \alpha\,\tau_x\big) \hat{S}_i\cdot\bm{\sigma}\Big] \Psi(\vr{r}_i)
= -\sum_{j\neq i} \lim_{\substack{E\to 0\\\alpha\to 1}} J_E(\vr{r}_i-\vr{r}_j)\, \hat{S}_j\cdot\bm{\sigma}\, \Psi(\vr{r}_j).
\end{gathered}
\end{equation}
We will now consider a ferromagnetic arrangement with spins pointing in the $z$ direction, $\vr{S}_i = S\hat{e}_z$, and project Eq.~(\ref{J}) to the decoupled Shiba states $|\tau_x+\rangle|\uparrow\rangle$ and $|\tau_x-\rangle|\downarrow\rangle$. The function $J_E(\vec{r})$ on the RHS is given by 
\begin{align}
J_E(\vr{r}) &= 
- \frac{\alpha}{2} \Big\{ \big[ I_1^-(\vr{r}) + I_1^+(\vr{r}) \big] (E\tau_0\sigma_0 + \Delta\tau_x\sigma_0)
- \big[ I_2^-(\vr{r}) - I_2^+(\vr{r}) \big] \tau_z\sigma_x
+ \big[ I_3^-(\vr{r}) - I_3^+(\vr{r}) \big] \tau_z\sigma_y \Big\} +g(\tau_z, \sigma_{x/y}, \tau_x\sigma_{x/y}) \nonumber
,
\end{align}
where $g(\tau_z, \sigma_{x/y}, \tau_x\sigma_{x/y})$ stand for terms proportional to $\tau_z\sigma_0, \tau_0\sigma_{x,y}$ and $ \tau_x\sigma_{x,y}$ that have only vanishing matrix elements between the low-energy basis states. In addition we have defined the functions
\begin{equation*}
\begin{split}
I_1^\pm(\vr{r}) &= \frac{1}{2\pi^2} \frac{\mathcal{N}_\pm}{\mathcal{N}} \intt d\varphi\! \intt d\xi\, \frac{e^{ik^\pm(\xi)\,r\cos\beta}}{\Delta^2 - E^2 + \xi^2} 
,\\
I_2^\pm(\vr{r}) &= \frac{1}{2\pi^2} \frac{\mathcal{N}_\pm}{\mathcal{N}} \intt d\varphi\! \intt d\xi\, \frac{e^{ik^\pm(\xi)\,r\cos\beta}\, \xi\sin\varphi}{\Delta^2 - E^2 + \xi^2}
,\\
I_3^\pm(\vr{r}) &= \frac{1}{2\pi^2} \frac{\mathcal{N}_\pm}{\mathcal{N}} \intt d\varphi\! \intt d\xi\, \frac{e^{ik^\pm(\xi)\,r\cos\beta}\, \xi\cos\varphi}{\Delta^2 - E^2 + \xi^2}.
\end{split}
\end{equation*}
Here $r=|\vr{r}|$, $\hat{r}=\vr{r}/r$ and $\beta$ is the angle between $\vr{r}$ and $\vr{k}$. We have defined $k^\pm(\xi) = k_F^\pm + \xi/(\hbar\tilde{v}_F)$, while $k_F^\pm = k_F \big( \sqrt{1 + \lambda\displaystyle{^2}} \mp \lambda \big)$ and $\tilde{v}_F = v_F \sqrt{1 + \lambda\displaystyle{^2}}$ are the Fermi wave vector and the Fermi velocity of the two helicity bands and $\lambda = \alpha_R/(\hbar v_F)$ is the dimensionless SOI strength. $\mathcal{N}_{\pm} = \mathcal{N} \Big[ 1 \mp \lambda\big/\sqrt{1 + \lambda\displaystyle{^2}} \Big]$ are the density of states of the helicity bands at the Fermi level.

Due to spherical symmetry, the integrals $I_1^\pm(\vr{r})$ can be evaluated straightforwardly. The integrals $I_2^\pm(\vr{r})$ and $I_3^\pm(\vr{r})$ which also depend on the direction of $\vr{r}$ have a general form
\begin{equation*}
\begin{gathered}
\intt\frac{d\vr{k}}{(2\pi)^2} e^{-\vr{k}\cdot\vr{r}} (\bm{\sigma}\times\hat{k}) = f(r)(\bm{\sigma}\times\hat{r}) 
= f(r)(\sigma_x\sin\varphi' - \sigma_y\cos\varphi'),
\end{gathered}
\end{equation*}
where $\vr{r} = r(\cos\varphi',\sin\varphi')$ and $f$ is a function of $r$. The function $f(r)$ can be fixed by evaluating a special case $\vec{r}=r\hat{e}_x$. Hence we have
\begin{equation*}
\begin{split}
&I_1^\pm(\vr{r}) = \frac{\mathcal{N}_\pm}{\mathcal{N}} \frac{1}{\sqrt{\Delta^2 - E^2}} \, \Re \Big[ J_0\big( k_F^\pm r +ir/\xi_E \big) +iH_0\big( k_F^\pm r +ir/\xi_E \big) \Big]
,\\
&I_2^\pm(\vr{r}) = i\, \frac{\mathcal{N}_\pm}{\mathcal{N}}\, \Re \Big[ iJ_{1}\big(k_F^\pm r +ir/\xi_E\big) + H_{-1}\big(k_F^\pm r +ir/\xi_E\big) \Big] \sin\varphi'
\equiv i\, I_0^\pm(r) \sin\varphi'
,\\
&I_3^\pm(\vr{r}) = i\, \frac{\mathcal{N}_\pm}{\mathcal{N}}\, \Re \Big[ iJ_{1}\big(k_F^\pm r +ir/\xi_E\big) + H_{-1}\big(k_F^\pm r +ir/\xi_E\big) \Big] \cos\varphi' 
\equiv i\, I_0^\pm(r) \cos\varphi'
,
\end{split}
\end{equation*}
where $J_n$ and $H_n$ are Bessel and Struve functions of order $n$, respectively, and $\xi_E = \hbar \tilde{v}_F/\sqrt{\Delta^2 - E^2}$. For large values of the argument these integrals can be given in asymptotic form as
\begin{equation*}
\begin{split}
&I_1^\pm(\vr{r}) \approx \frac{\mathcal{N}_\pm}{\mathcal{N}} \frac{1}{\sqrt{\Delta^2 - E^2}} \, \sqrt{\frac{2/\pi}{k_F^\pm r}} \cos\big(k_F^\pm r -\textstyle{\frac{\pi}{4}}\big)\, e^{-r/\xi_E}
,\\
&I_0^\pm(\vr{r}) \approx -\frac{\mathcal{N}_\pm}{\mathcal{N}} \Bigg[ \sqrt{\frac{2/\pi}{k_F^\pm r}} \sin\big(k_F^\pm r -\textstyle{\frac{3\pi}{4}}\big)\, e^{-r/\xi_E} + \frac{2/\pi}{(k_F^\pm r)^2} \Bigg]
.
\end{split} \tag{A.4}
\end{equation*}
The above approximations receive corrections of the order of $\mathcal{O}\big((k_F^\pm r)^{-3}\big)$.

Having evaluated the integrals, Eq.~(\ref{J}) projected to the two Shiba states becomes
\begin{equation*}
\begin{gathered}
\sum_k \begin{pmatrix} h_{jk} & \Delta_{jk} \\ \Delta_{kj}^* & -h_{jk} \end{pmatrix} 
\begin{pmatrix} u(\vr{r}_k) \\ v(\vr{r}_k) \end{pmatrix}
= E \begin{pmatrix} u(\vr{r}_j) \\ v(\vr{r}_j) \end{pmatrix}
\end{gathered}
\end{equation*}
where $u(\vr{r}_j) \equiv \langle\tau_x{+}|\langle\uparrow| \Psi(\vr{r}_j)\rangle$ and $v(\vr{r}_j) \equiv \langle\tau_x{-}|\langle\downarrow| \Psi(\vr{r}_j)\rangle$ are the amplitudes for the decoupled states with energies $\pm\epsilon_0$  at position $\vr{r}_j$ and 
\begin{equation*}
\begin{split}
h_{ij} &= 
\Bigg\{\!\!\begin{array}{cl}
\epsilon_0 & i=j
\\
\displaystyle
-\frac{\Delta^2}{2} \lim_{E\rightarrow 0} \big[ I_1^-(r_{ij}) + I_1^+(r_{ij}) \big] & i\neq j
\end{array} ,
\\
\Delta_{ij} &= 
\Bigg\{\!\!\begin{array}{cl}
0 & i=j
\\
\displaystyle
\frac{\Delta}{2} \lim_{E\rightarrow 0} \big[ I_0^+({r}_{ij}) - I_0^-({r}_{ij}) \big] \frac{x_{ij} - iy_{ij}}{r_{ij}} & i\neq j 
\end{array} ,
\end{split}
\end{equation*}
and $\epsilon_0 = \Delta(1-\alpha)$ is the energy of an isolated impurity state. Evaluating the limit $\tilde{I}_{0/1}^{\pm}=\lim_{E\rightarrow 0} I_{0/1}^{\pm}$ we obtain Eq.~(3) in the main text.
\end{document}